\documentclass[useAMS,usenatbib]{mn2e}

\usepackage{graphicx}  
\usepackage{ amssymb }
\usepackage{journals}
\usepackage[dvips]{color}
\newcommand{\teff}{$T_{\rm eff}$} 
\newcommand{\logg}{$\log g$} 
\newcommand{\kms}{km s$^{-1}$}
\newcommand{\vt}{$\xi_t$} 
\newcommand{\fei}{Fe\,{\sc i}}
\newcommand{\feii}{Fe\,{\sc ii}}
\newcommand{\scii}{Sc\,{\sc ii}}
\newcommand{\oi}{O\,{\sc i}}
\newcommand{\nai}{Na\,{\sc i}}
\newcommand{\vi}{V\,{\sc i}}
\newcommand{\mni}{Mn\,{\sc i}}
\newcommand{\coi}{Co\,{\sc i}}
\newcommand{\yii}{Y\,{\sc ii}}
\newcommand{\sri}{Sr\,{\sc i}}
\newcommand{\rbi}{Rb\,{\sc i}}
\newcommand{\pbi}{Pb\,{\sc i}}
\newcommand{\baii}{Ba\,{\sc ii}}
\newcommand{\laii}{La\,{\sc ii}}
\newcommand{\prii}{Pr\,{\sc ii}}
\newcommand{\euii}{Eu\,{\sc ii}}
\newcommand{\tii}{Ti\,{\sc i}}
\newcommand{\tiii}{Ti\,{\sc ii}}

\newcommand\solmass{M$_{\odot}$}
\newcommand\simgt{\lower.3ex\hbox{\gtsima}}

\title[Chemical abundances in NGC 6266]{Chemical abundances in bright giants of
the globular cluster M62 (NGC 6266)\thanks{Based in part on data collected at
Subaru Telescope, which is operated by the National Astronomical Observatory of
Japan. This paper includes data gathered with the 6.5 meter Magellan
Telescopes located at Las Campanas Observatory, Chile.}} 
\author[D.\ Yong et al.]
{David Yong$^{1,2}$\thanks{E-mail: david.yong@anu.edu.au}, 
Alan Alves Brito$^1$, 
Gary S.\ Da Costa$^1$, 
Javier Alonso-Garc{\'{\i}}a$^{3,4}$, \newauthor
Amanda I.\ Karakas$^1$, 
Marco Pignatari$^5$, 
Ian U.\ Roederer$^6$,  
Wako Aoki$^7$, \newauthor 
Cherie K.\ Fishlock$^1$, 
Frank Grundahl$^8$ and 
John E.\ Norris$^1$. \\ 
$^{1}$Research School of Astronomy and Astrophysics, Australian
National University, Canberra, ACT 2611, Australia\\ 
$^{2}$Stromlo Fellow\\
$^{3}$Instituto de Astrof\'isica, Facultad de F\'isica, Pontificia Universidad
Cat\'olica de Chile, Av. Vicu\~na Mackenna 4860, 782-0436 Macul, Santiago,
Chile\\ 
$^{4}$The Milky Way Millennium Nucleus, Av. Vicu\~{n}a Mackenna 4860, 782-0436
Macul, Santiago, Chile\\ 
$^{5}$Department of Physics, University of Basel, Klingelbergstrasse 82,
CH-4056 Basel, Switzerland\\ 
$^{6}$Department of Astronomy, University of Michigan, 500 Church Street, Ann
Arbor, MI 48109, USA\\ 
$^{7}$National Astronomical Observatory, Mitaka, Tokyo 181-8588,
Japan\\
$^{8}$Stellar Astrophysics Centre, Department of Physics and 
Astronomy, Aarhus University, Ny Munkegade 120, DK-8000 Aarhus C, Denmark\\
}
\begin{document}

\date{Accepted 2014 January 15.  Received 2014 January 14; in original form 2013 November 26}

\pagerange{\pageref{firstpage}--\pageref{lastpage}} \pubyear{2014}

\maketitle

\label{firstpage}

\begin{abstract}
With the exception of Terzan 5, all the Galactic globular clusters that possess
significant metallicity spreads, such as $\omega$ Cen and M22, are
preferentially the more luminous clusters with extended horizontal branches.
Here we present radial velocities and chemical abundances for seven bright
giants in the globular cluster M62, a previously little-studied cluster.  With
$M_V$ = $-$9.18, M62 is the ninth most luminous Galactic globular cluster and
has an extended horizontal branch. Within our sample, we find (i) no evidence
for a dispersion in metallicity, [Fe/H], beyond the measurement uncertainties,
(ii) star-to-star abundance variations for C, O, Na and Al with the usual
correlations between these elements as seen in other globular clusters, and
(iii) a global enrichment for the elements Zr, Ba and La at the level [X/Fe]
$\simeq$ +0.4 dex. For elements heavier than La, the abundance ratios are
consistent with the scaled-solar $r$-process distribution. Below La, the
abundances are anomalous when compared to the scaled-solar $s$-process or
$r$-process distributions. For these elements, the abundance signature in M62
is in agreement with predictions of the $s$-process from fast-rotating massive
stars, although the high [Rb/Y] ratio we measure may be a challenge to this
scenario. 
\end{abstract}


\section{INTRODUCTION}

A small, but growing, number of Galactic globular clusters exhibit a
star-to-star variation in the relative abundances of the heavy elements.
$\omega$ Centauri is the most well-known member of this group with its stars
spanning a range in metallicity\footnote{Here and throughout the paper, iron is
the canonical measure of metallicity and we adopt the standard spectroscopic
notation [Fe/H] = $\log_{10}(\rm N_{\rm Fe}/N_{\rm H})_\star - \log_{10}(N_{\rm
Fe}/N_{\rm H})_\odot$} from [Fe/H] $\simeq$ $-$2.0 to [Fe/H] $\simeq$ $-$0.5;
for certain elements, the abundance ratios with respect to iron, [X/Fe],
exhibit considerable variations with metallicity (e.g.,
\citealt{norris95,smith00,johnson10,dorazi11}). As the most massive Galactic
globular cluster and in light of its complex chemical enrichment history,
$\omega$ Cen is regarded to be the core of a disrupted dwarf galaxy (e.g.,
\citealt{freeman93,bekki03,bekki06}). 

M54 is the second most massive Galactic globular cluster. It displays a key
characteristic shared by $\omega$ Centauri but absent in the majority of
clusters, namely, a metallicity dispersion \citep{carretta10}, albeit of
smaller amplitude than that found in $\omega$ Centauri. That M54 is the nuclear
star cluster of the Sagittarius dwarf spheroidal galaxy lends further support
to the hypothesis that massive globular clusters with metallicity variations
may be the nuclei of disrupted dwarf galaxies (although \citealt{bellazzini08}
argue that the nucleus of Sgr likely formed independently of M54). When
considering the ensemble of globular clusters with unusual chemical abundance
ratios\footnote{In addition to $\omega$ Cen and M54, the other clusters include
M15 \citep{sneden00,otsuki06,sobeck11,worley13}, NGC 362 \citep{carretta13},
NGC 1851 \citep{yong081851,villanova10,carretta11}, NGC 2419
\citep{cohen10,cohen11,cohen12,mucciarelli12}, NGC 3201 \citep{simmerer13}, NGC
5824 \citep{saviane12}, NGC 6656
\citep{dacosta09,dacosta11,marino09,marino11,roederer11,alvesbrito12}, NGC 7089
\citep{lardo12,lardo13} and Terzan 5 \citep{ferraro09,origlia11,origlia13}}
(dispersions in metallicity and/or neutron-capture elements), it is clear that
these objects are preferentially the more luminous, and therefore the more
massive, clusters. 

\citet{lee07} considered globular clusters with an extended horizontal branch
(i.e., a horizontal branch with extremely blue stars) which likely signifies
the presence of He abundance variations \citep{dantona02}. They argued that not
only are clusters with an extended horizontal branch preferentially more
massive, but also when considered together the clusters have different
kinematics from other groupings of the Galactic globular cluster population.
The sample of clusters with extended horizontal branches include $\omega$ Cen,
M54 and M22, all of which exhibit metallicity variations.  \citet{lee07} showed
that such clusters had kinematics consistent with accretion and thus likely
have a different origin to the bulk of the cluster population. 

The identification of additional globular clusters with peculiar chemical
abundance patterns (e.g., star-to-star variations in metallicity) would advance
our understanding of the range and relative frequency of the abundance spread
phenomenon in the Galactic globular cluster system. Additionally, quantifying
the number of accreted systems has important consequences for understanding the
formation and evolution of our Galaxy \citep{freeman02}. To this end, the most
obvious objects to study are the most massive globular clusters with extended
horizontal branches.  M62 (NGC 6266) is the ninth most luminous cluster
\citep{harris96} and has an extended blue horizontal branch.  In the F439W $-$
F555W versus F555W colour magnitude diagram presented by \citet{piotto02}, the
horizontal branch extends down in luminosity to the main sequence turn-off. The
extended horizontal branch is also readily identified in dereddened $V$ versus
$B-V$ and $V$ versus $V-I$ diagrams \citep{alonsogarcia11,alonsogarcia12}. M62
ranks fifth in the number of millisecond pulsars \citep{cocozza08}, and there
may be a connection between multiple stellar populations and the numbers of
millisecond pulsars \citep{lanzoni10}.  \citet{contreras10} find that M62 could
contain the most RR-Lyrae of any Galactic globular cluster and is a typical
Oosterhoff type I system. This cluster lies in the vicinity of the bulge,
although proper-motion measurements indicate that it probably belongs to the
thick disk \citep{dinescu03}.  M62 is highly reddened, $E(B-V$) = 0.47
\citep{harris96}, and it is severely affected by differential reddening,
$\Delta E(B-V)$ $\sim$ 0.25 \citep{alonsogarcia12}. 

The goal of the present paper is to conduct a detailed chemical abundance
analysis of M62 based on high-resolution, high signal-to-noise ratio spectra.
To our knowledge this is the first such analysis of this cluster and this work
therefore represents an essential step towards completing the census and
characterisation of the Galaxy's most massive globular clusters. 

\section{OBSERVATIONS AND ANALYSIS}
\label{sec:obs}

\subsection{Target selection and spectroscopic observations} 

Six targets were selected from \citet{rutledge97} (see Table \ref{tab:param}).
All were high probability members based on the published radial velocities and
were observed with the High Dispersion Spectrograph (HDS, \citealt{noguchi02})
on the Subaru Telescope on 2013 July 17. The exposure times ranged from 16 to
40 minutes per program star. We used the StdYb setting and the 0\farcs8 slit
which resulted in a wavelength coverage from $\sim$4100\AA\ to $\sim$6800\AA\
at a spectral resolution of $R = 45,000$. There were approximately four pixels
per resolution element.  Spectra of a telluric standard (HD 163955) and a
radial velocity standard (HD 182572) were also taken during this observing run. 

Two additional targets were selected from Str\"omgren photometry obtained by
J.A-G and observed using the MIKE spectrograph \citep{bernstein03} at the
Magellan-II Telescope on 2012 June 26. The exposure times were 1 hour for each
star.  The 0\farcs7 slit was employed providing a spectral resolution of $R$ =
35,000 (blue CCD) and $R$ = 27,000 (red CCD) with a wavelength coverage from
$\sim$3800\AA\ to $\sim$9000\AA. There were approximately four pixels per
resolution element. A telluric standard (HD 170642) was also observed. 

\begin{figure}
\centering
      \includegraphics[width=.99\hsize]{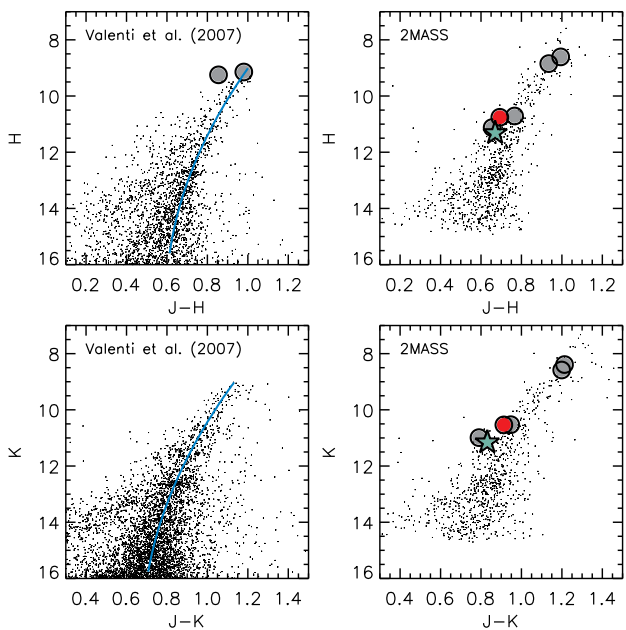} 
      \caption{$JHK$ colour-magnitude diagrams for M62 from \citet{valenti07}
(left panels) and 2MASS \citet{2mass} (right panels; taking only objects within
a radius of 3.5' and ``AAA'' photometric quality flags). The aqua star 
represents the radial velocity non-member, star A195, and the red circle 
represents star A124, the possible non-member based on chemical abundances. All
other program stars are shown as grey circles. In the left panels, the solid
line represents the mean RGB ridge line from \citet{valenti07}.} 
         \label{fig:cmd}
\end{figure}

In Figure \ref{fig:cmd}, we show the locations of our targets in $JHK$
colour-magnitude diagrams (CMD). The photometry is from \citet{valenti07} and
2MASS \citep{2mass}. 

\begin{table*}
 \centering
 \begin{minipage}{180mm}
  \caption{Program stars, radial velocities and stellar parameters.}
  \label{tab:param} 
  \begin{tabular}{@{}lccrrrrcccccc@{}}
  \hline
	Name\footnote{AXXX names are from \citet{alcaino78} while c53 and d10
are from Str\"omgren photometry taken by J.A-G.}  & 
	RA 2000 & 
	Dec.\ 2000 & 
	$J$\footnote{Photometry for the AXXX stars is from 2MASS \citep{2mass}.
Photometry for c53 and d10 are from \citet{valenti07}.} & 
	$H^b$ & 
	$K^b$ & 
	RV &
	$\sigma$RV\footnote{The uncertainties are the standard error of the mean.} & 
	\teff\  & 
	\logg\ & 
	\vt\ & 
	[m/H] & 
	[Fe/H] \\ 
	 & 
	 & 
	 & 
	 & 
	 & 
	 & 
	(\kms) & 
	(\kms) & 
	(K) & 
	(cm s$^{-2}$) & 
	(\kms) & 
	dex & 
	dex \\
	(1) & 
	(2) &
	(3) &
	(4) & 
	(5) & 
	(6) & 
	(7) & 
	(8) & 
	(9) & 
	(10) & 
	(11) & 
	(12) & 
	(13) \\ 
  \hline
\multicolumn{13}{c}{Subaru Telescope Observations 2013 07 17} \\ 
\hline 
   A5 & 17 01 16.21 & $-$30 03 31.81 &  9.782 &  8.847 &   8.583 & $-$75.0 & 0.2 &   3950 &   0.50 &   1.79 & $-$1.10 & $-$1.12 \\
   A7 & 17 01 17.50 & $-$30 03 20.60 &  9.602 &  8.608 &   8.389 & $-$89.3 & 0.5 &   3925 &   0.20 &   1.89 & $-$1.10 & $-$1.15 \\
  A14 & 17 01 17.29 & $-$30 03 58.69 & 11.476 & 10.710 &  10.529 & $-$54.2 & 0.2 &   4400 &   1.40 &   1.53 & $-$1.10 & $-$1.10 \\
  A27 & 17 01 25.19 & $-$30 04 09.85 & 11.779 & 11.124 &  10.989 & $-$69.9 & 0.4 &   4625 &   1.50 &   1.76 & $-$1.20 & $-$1.24 \\
 A124 & 17 01 00.28 & $-$30 07 38.28 & 11.449 & 10.756 &  10.536 & $-$68.2 & 0.3 &   4450 &   1.00 &   1.64 & $-$1.20 & $-$1.19 \\
 A195 & 17 01 12.63 & $-$30 03 40.78 & 11.988 & 11.318 &  11.157 &  $-$0.1 & 0.4 & \ldots & \ldots & \ldots &  \ldots &  \ldots \\
\hline
\multicolumn{13}{c}{Magellan Telescope Observations 2012 06 26} \\ 
\hline
  c53 & 17 01 15.20 & $-$30 08 36.50 & 10.101 &  9.246 &  \ldots & $-$73.6 & 0.2 &   4450 &   1.40 &   1.78 & $-$1.10 & $-$1.11 \\
  d10 & 17 01 16.80 & $-$30 08 18.00 & 10.125 &  9.145 &  \ldots & $-$59.0 & 0.2 &   4175 &   0.90 &   1.90 & $-$1.10 & $-$1.15 \\
\hline
\end{tabular}
\end{minipage}
\end{table*}

\subsection{Data reduction and analysis}

\begin{figure}
\centering
      \includegraphics[width=.99\hsize]{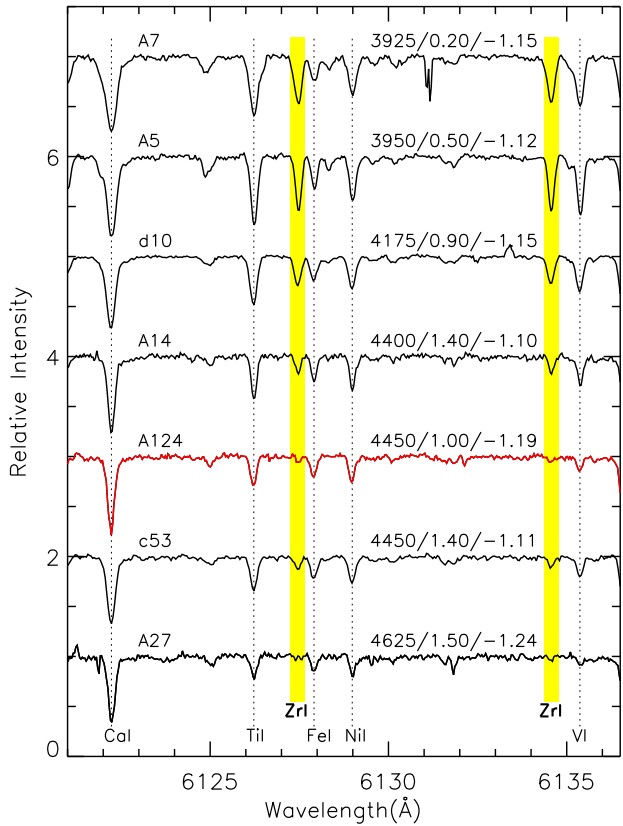} 
      \caption{A portion of the spectra for the program stars. The yellow
region highlights two Zr lines used in the analysis. The positions of other
atomic lines and the stellar parameters (\teff/\logg/[Fe/H]) are included. The
red spectrum is star A124, a possible non-member based on the chemical
abundances.} 
         \label{fig:spec}
\end{figure}

The Subaru spectra were reduced using IRAF\footnote{IRAF is distributed by the
National Optical Astronomy Observatories, which are operated by the Association
of Universities for Research in Astronomy, Inc., under cooperative agreement
with the National Science Foundation.}. The Magellan spectra were reduced using
a combination of IRAF and the mtools
package\footnote{www.lco.cl/telescopes-information/magellan/instruments/mike/iraf-tools/iraf-mtools-package}.
In both cases, the approach was similar to that described in \citet{yong06}. 
In Figure \ref{fig:spec}, we plot a region of the spectra for the program
stars. The telluric standards were used to correct regions affected by telluric
absorption. Star A195 had a signal-to-noise ratio S/N = 40 per pixel
near 6000\AA. Excluding this star, the S/N ranged from S/N = 65 per pixel
(stars A14 and A27) to S/N = 110 per pixel (star d10). 

For the program stars observed with Subaru, heliocentric radial velocities were
measured by cross-correlating each star against the radial velocity standard,
HD 182572, for which we adopted a
value\footnote{obswww.unige.ch/$\sim$udry/std/stdnew.dat} of $-$100.35 \kms.
For the program stars observed with Magellan, the radial velocities were
determined by measuring the positions of a large number (several hundred) of
atomic absorption features. The radial velocities are presented in Table
\ref{tab:param} and we note that star A195 is not a radial velocity member. 

Equivalent widths (EW) were measured using routines in IRAF and DAOSPEC
\citep{stetson08}. We note that every line measured using DAOSPEC was also
measured, and visually inspected, using IRAF. For all stars, there was good
agreement between the two sets of measurements; for a given star, the average
difference in EWs between IRAF and DAOSPEC was 0.1 m\AA\ with a dispersion of 3
m\AA. The line list and EW measurements (average of IRAF and DAOSPEC) are
presented in Table \ref{tab:ew}. 

\begin{table*}
 \centering
 \begin{minipage}{180mm}
  \caption{Line list and equivalent widths.}
  \label{tab:ew} 
  \begin{tabular}{@{}cccrrrrrrrrc@{}}
  \hline
         Wavelength & 
         Species\footnote{The digits to the left of the decimal point are the
atomic number. The digit to the right of the decimal point is the ionization
state (‘0’ = neutral, ‘1’ = singly ionized).}  & 
         L.E.P & 
         $\log gf$ & 
         A5 & 
         A7 & 
         A14 & 
         A27 & 
         A124 & 
         c53 & 
         d10 & 
         Source\footnote{A = $\log gf$ values taken from \citet{yong05} where the references include 
\citet{denhartog03}, \citet{ivans01}, \citet{kurucz95}, 
\citet{prochaska00}, \citet{ramirez02}; 
B = \citet{gratton03};
C = Oxford group including 
\citet{blackwell79feb,blackwell79fea,blackwell80fea,blackwell86fea,blackwell95fea}; 
D = \citet{biemont91};
E1 = \citet{fuhr09}, 
E2 = \citet{roederer12c},
E3 = \citet{biemont11},
E4 = \citet{biemont81},
E5 = \citet{lawler01a}, 
E6 = \citet{lawler09},
E7 = \citet{li07}, 
E8 = \citet{denhartog03}, 
E9 = \citet{lawler06}, 
E10 = \citet{lawler01b},
E11 = \citet{roederer12b};
F = \citet{lambert76}} \\
         \AA\ & 
          & 
         eV & 
          & 
         m\AA & 
         m\AA & 
         m\AA & 
         m\AA & 
         m\AA & 
         m\AA & 
         m\AA & 
          \\ 
         (1) & 
         (2) &
         (3) &
         (4) & 
         (5) & 
         (6) & 
         (7) & 
         (8) & 
         (9) & 
         (10) & 
         (11) & 
         (12) \\ 
\hline
  6300.31 &   8.0 &   0.00 &  $-$9.75 &    33.0 &    48.9 & $<$15.0 &    67.0 &    48.8 & $<$15.0 &    60.2 &          B \\ 
  6363.78 &   8.0 &   0.02 & $-$10.25 &  \ldots &    17.0 &  \ldots &    21.2 &    22.1 &  \ldots &    34.4 &          A \\ 
  7771.95 &   8.0 &   9.15 &     0.35 &  \ldots &  \ldots &  \ldots &  \ldots &  \ldots &  \ldots &    12.8 &          B \\ 
  7774.18 &   8.0 &   9.15 &     0.21 &  \ldots &  \ldots &  \ldots &  \ldots &  \ldots &  \ldots &    10.1 &          B \\ 
  5682.65 &  11.0 &   2.10 &  $-$0.67 &  \ldots &  \ldots &   114.8 &    51.2 &  \ldots &   110.3 &    98.0 &          B \\ 
\hline 
\end{tabular}
\end{minipage}
This table is published in its entirety in the electronic edition of the paper.
A portion is shown here for guidance regarding its form and content. 
\end{table*}

The stellar parameters were determined using a traditional spectroscopic
approach. One dimensional local thermodynamic equilibrium (LTE) model
atmospheres were taken from the \citet{castelli03} alpha enhanced,
[$\alpha$/Fe] = +0.4, grid. To produce particular models we used the
interpolation software tested in \citet{allende04}. Chemical abundances were
generated using the LTE stellar line analysis program MOOG
\citep{moog,sobeck11}. The effective temperature, \teff, was obtained when
there was no trend between the abundance from \fei\ lines and the lower
excitation potential. The surface gravity, \logg, was established by forcing
the iron abundance derived from \fei\ and \feii\ lines to be equal. The
microturbulent velocity, \vt, was adjusted until the abundance from \fei\ lines
exhibited no trend with reduced equivalent width, $\log ({\rm EW}/\lambda)$.
Finally, we required the derived metallicity to be within 0.1 dex of the value
adopted in the model atmosphere. The final stellar parameters (see Table
\ref{tab:param}) were established only when all conditions above were
simultaneously satisfied.  The uncertainties in \teff, \logg\ and \vt\ are
estimated to be 50 K, 0.20 dex and 0.20 \kms, respectively.  We do not present
results for star A195 since it is not a radial velocity member and our
preliminary analysis indicated that it is a super-solar metallicity clump giant
(\teff\ = 4750, \logg\ = 3.1, [Fe/H] = +0.17). 

\begin{figure}
\centering
      \includegraphics[width=.90\hsize]{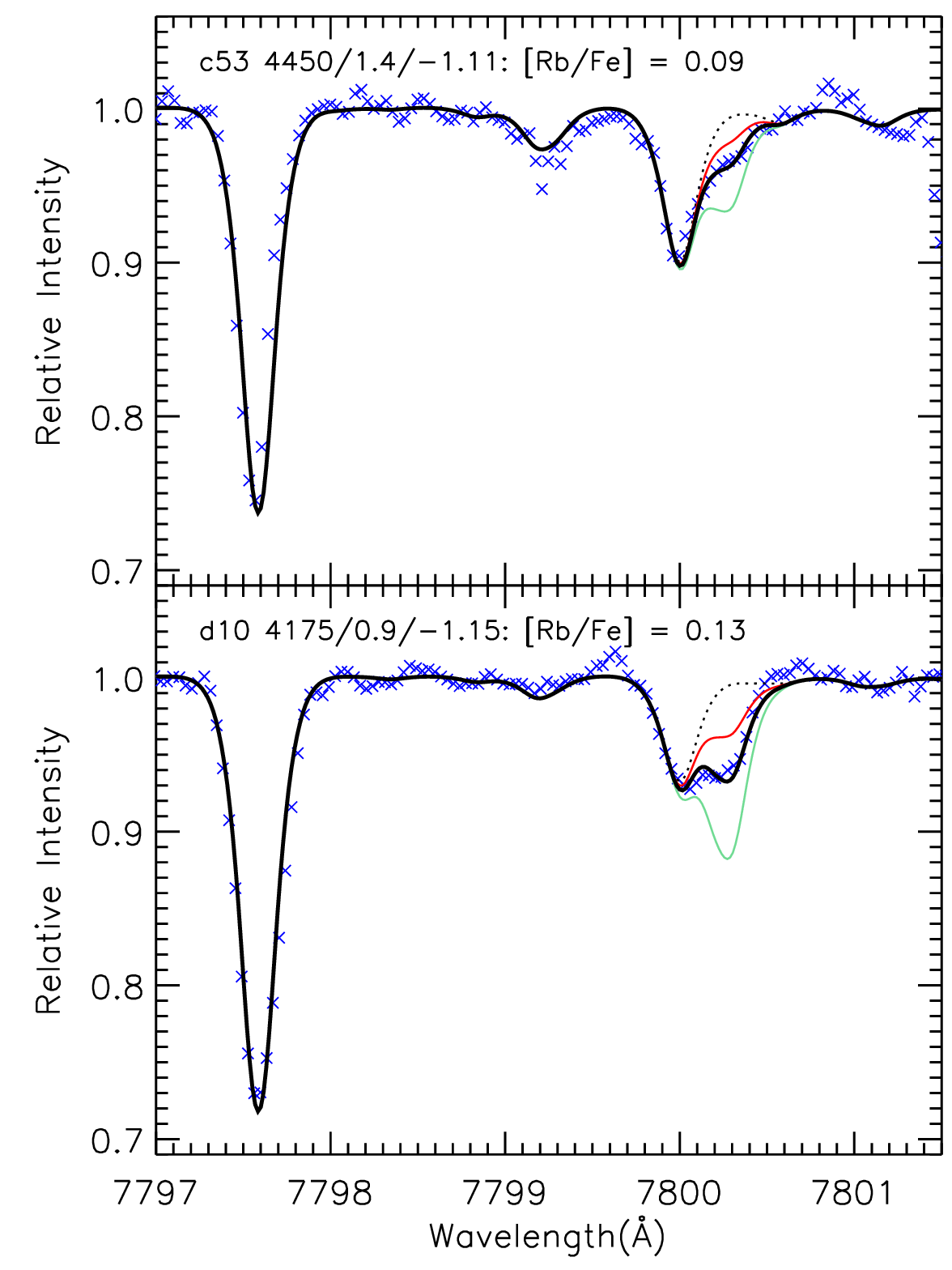} 
      \caption{Observed and synthetic spectra for stars c53 (upper) and d10
(lower) near the 7800\AA\ Rb feature. The thick black line represents the best
fit, thin red and green lines are $\Delta$[Rb/Fe] = $\pm$ 0.3 dex from the line
of best fit and the dotted line has no Rb. The stellar parameters
(\teff/\logg/[Fe/H]) and [Rb/Fe] abundance are included in each panel.} 
         \label{fig:rb}
\end{figure}

\begin{figure}
\centering
      \includegraphics[width=.90\hsize]{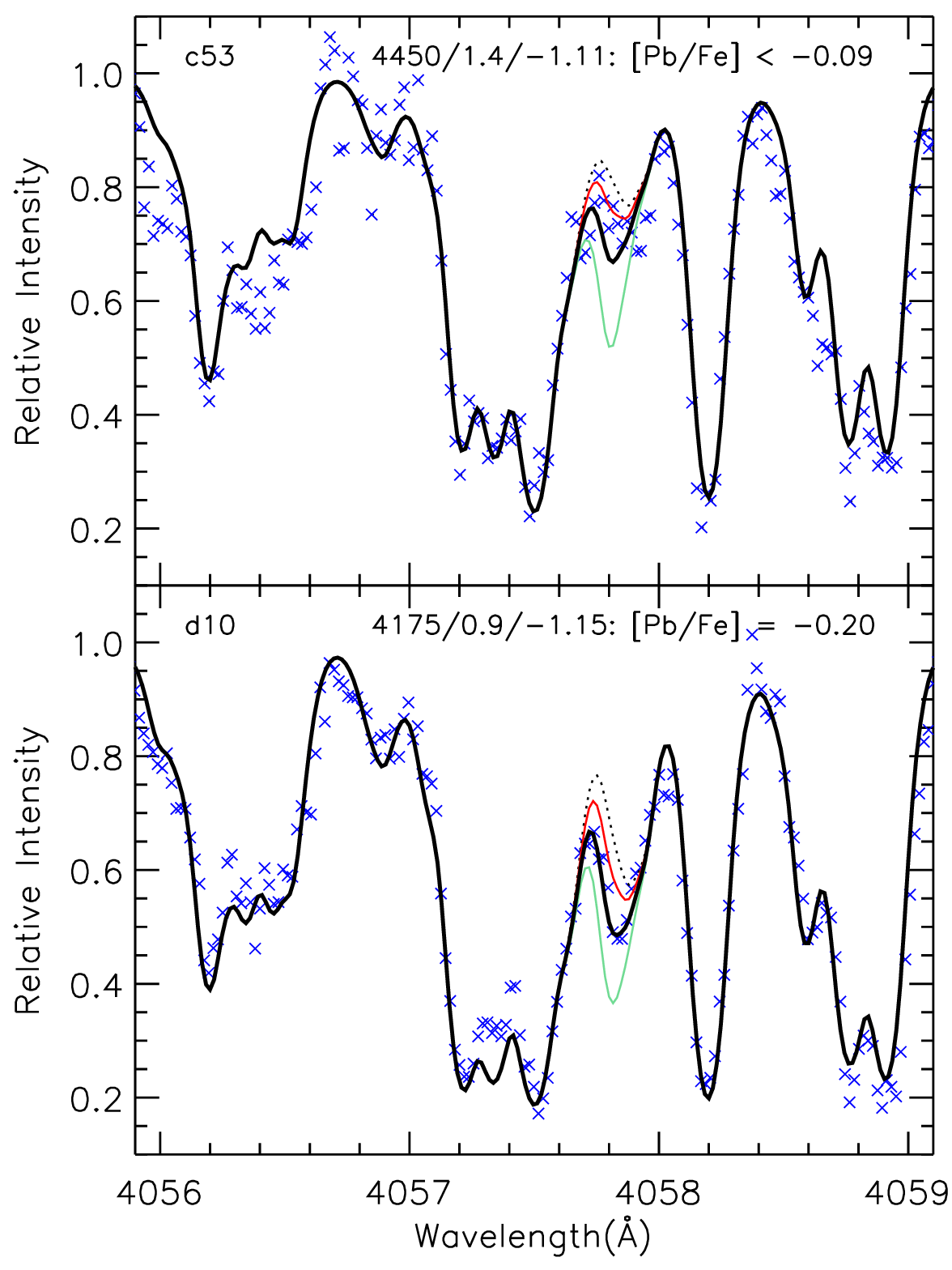} 
      \caption{Observed and synthetic spectra for stars c53 (upper) and d10
(lower) near the 4058\AA\ Pb feature. The thick black line represents the best
fit (d10) or abundance limit (c53), thin red and green lines are
$\Delta$[Pb/Fe] = $\pm$ 0.5 dex from the line of best fit and the dotted line
has no Pb. The stellar parameters (\teff/\logg/[Fe/H]) and [Pb/Fe] abundance
are included in each panel.} 
         \label{fig:pb}
\end{figure}

Chemical abundances were determined using the final model atmosphere, measured
EW and MOOG. For Cu and the neutron-capture elements, abundances were
determined via spectrum synthesis. Lines affected by hyperfine splitting (HFS)
and/or isotope shifts (e.g., \scii, \vi, \mni, \coi, \rbi, \baii, \laii,
\prii, \euii\ and \pbi) were treated appropriately using the HFS data from
\citet{kurucz95} or other sources as noted in Table \ref{tab:ew}. For Rb and
Pb, the wavelength coverage necessary to measure these elements was only
obtained in the two stars observed with Magellan (see Figures \ref{fig:rb} and
\ref{fig:pb}). C abundances were determined from spectrum synthesis of the
4300\AA\ CH molecular lines using the CH line list compiled by B.\ Plez et al.\
(2009, private communication). In our analysis, the dissociation energy for CH
was 3.465 eV. The \citet{asplund09} solar abundances were adopted and the
chemical abundances are presented in Table \ref{tab:abun}. 

\begin{table}
 \centering
 \begin{minipage}{90mm}
  \caption{Chemical abundances for the program stars.}
  \label{tab:abun} 
  \begin{tabular}{@{}lcrcrc@{}}
  \hline
	Name & 
	A(X) & 
	N$_{\rm lines}$ & 
	s.e.$_{\log\epsilon}$ & 
	[X/Fe] & 
	Total Error \\
\hline 
\hline
     \multicolumn{6}{c}{C from CH}   \\
\hline
A5     &    6.46 & synth &  \ldots & $-$0.85 &    0.21 \\
A7     &    6.31 & synth &  \ldots & $-$0.97 &    0.23 \\
A14    &    6.41 & synth &  \ldots & $-$0.92 &    0.21 \\
A27    &    6.76 & synth &  \ldots & $-$0.43 &    0.22 \\
A124   &    6.76 & synth &  \ldots & $-$0.48 &    0.23 \\
c53    &    6.71 & synth &  \ldots & $-$0.61 &    0.22 \\
d10    &    7.01 & synth &  \ldots & $-$0.27 &    0.22 \\
\hline
     \multicolumn{6}{c}{\oi}   \\
\hline
A5     &    7.79 & 2      &    0.07 &    0.30 &    0.18 \\
A7     &    7.79 & 2      &    0.04 &    0.39 &    0.20 \\
A14    &    7.68 & 1      &  \ldots & $<$0.16 &  \ldots \\
A27    &    8.40 & 2      &    0.09 &    1.05 &    0.18 \\
A124   &    8.08 & 2      &    0.03 &    0.68 &    0.18 \\
c53    &    7.66 & 1      &  \ldots & $<$0.17 &  \ldots \\
d10    &    8.31 & 4      &    0.04 &    0.89 &    0.14 \\
\hline
     \multicolumn{6}{c}{\nai}   \\ 
\hline
A5     &    5.67 & 2      &    0.04 &    0.64 &    0.18 \\
A7     &    5.63 & 2      &    0.00 &    0.68 &    0.17 \\
A14    &    5.79 & 3      &    0.04 &    0.72 &    0.13 \\
A27    &    4.87 & 3      &    0.05 & $-$0.04 &    0.13 \\
A124   &    4.86 & 2      &    0.01 & $-$0.09 &    0.15 \\
c53    &    5.59 & 3      &    0.06 &    0.55 &    0.12 \\
d10    &    5.18 & 4      &    0.04 &    0.21 &    0.12 \\
\hline
\end{tabular}
\end{minipage}
This table is published in its entirety in the electronic edition of the paper.
A portion is shown here for guidance regarding its form and content. 
\end{table}

The uncertainties in the chemical abundances were obtained in the following
manner. We repeated the analysis and varied the stellar parameters, one at a
time, by their uncertainties. We also considered the uncertainty of the
metallicity used to generate the model atmosphere, [M/H], and adjusted this
value by +0.2 dex. The systematic error was determined by adding these four
error terms in quadrature, and we assume these values are symmetric for
positive and negative changes. Following \citet{norris10}, the random error
(s.e.$_{\log\epsilon}$) was obtained by taking max(s.e.$_{\log \epsilon}$,
0.20/$\sqrt{N_{\rm lines}}$).  The second term is what would be expected for a
set of $N_{\rm lines}$ with a dispersion of 0.20 dex (a conservative value
based on the abundance dispersion exhibited by \fei\ lines). By adding the
systematic and random errors, in quadrature, we obtain the total error which is
presented in Table \ref{tab:abun}. 

\section{RESULTS}

\subsection{Cluster membership and radial velocities} 

With the exception of star A195, all program stars have radial velocities
consistent with cluster membership. According to the \citet{harris96}
catalogue, M62 has a heliocentric radial velocity of $-$70.1 $\pm$ 1.4 \kms\
and a central velocity dispersion of 14.3 $\pm$ 0.4 \kms. For our seven program
stars, we find an average heliocentric radial velocity of $-$69.9 $\pm$ 4.3
\kms\ ($\sigma$ = 11.4 \kms). 

As we shall discuss in the following subsection, star A124 has low abundance
ratios for the neutron-capture elements compared to the other program stars and
so the possibility exists that this object may not be a cluster member. With a
radial velocity of $-$68.2 $\pm$ 0.3 \kms, excluding this star would not
substantially change the mean radial velocity and velocity dispersion we
measure for this cluster. 

To investigate the likelihood of observing field stars in the vicinity of M62
with stellar parameters (\teff, \logg\ and [Fe/H]) and radial velocities
comparable to the program stars, we make use of the Besan\c{c}on model
\citep{robin03}. First, we consider all stars within a one square degree field
centered on M62. Secondly, we restricted the sample to lie near the RGB in the
$K$ versus $J-K$ CMD; specifically, the left edge was defined as the line from
($J-K$,$K$) = (0.55,12.0) to ($J-K$,$K$) = (1.1,8.0) and the right edge as the
line from ($J-K$,$K$) = (1.05,12.0) to ($J-K$,$K$) = (1.4,8.0). We find 13281
such stars from the Besan\c{c}on model. Thirdly, of these 13281 stars we
counted the number that satisfied the following constraints: ($i$) $-$95 $\le$
RV $\le$ $-$45 \kms\ and ($ii$) $-$1.35 $\le$ [Fe/H] $\le$ $-$1.00 dex. And
finally, we counted the numbers of stars that lay in a particular region in the
\teff-\logg\ plane, specifically, the area is bounded at the left edge by the
line from (\teff,\logg) = (4925,2.0) to (\teff,\logg) = (4225,0.0), at the
right edge by the line from (\teff,\logg) = (4200,2.0) to (\teff,\logg) =
(3625,0.0) both with 0.0 $\le$ \logg\ $\le$ 2.0.  We found 490 stars in the
Besan\c{c}on model that satisfied all criteria and therefore estimate that
given a sample of stars occupying similar locations in $JHK$ colour-magnitude
diagrams as the program stars, the probability of observing a field star with
stellar parameters and a radial velocity consistent with the program stars is
roughly 4\%. In our sample of seven stars, having one field star with stellar
parameters and a radial velocity consistent with the cluster sample is not
unreasonable given the small number statistics.  Accurate proper-motion and
parallax measurements from GAIA would establish cluster membership, or
otherwise, for this object. 

\subsection{Chemical abundances} 

For the program stars, we measure a mean metallicity of [Fe/H] = $-$1.15 $\pm$
0.02 ($\sigma$ = 0.05). Our metallicity is consistent with previous
measurements based primarily on low-resolution spectroscopic measurements:
[Fe/H] = $-$1.28 $\pm$ 0.15 \citep{zinn84}; [Fe/H] = $-$1.02 $\pm$ 0.04
\citep{carretta97}; [Fe/H] = $-$1.19 \citep{kraft03} when using MARCS model
atmospheres \citep{gustafsson75}; [Fe/H] = $-$1.18 $\pm$ 0.07 \citep{Ca09}.
The dispersion in [Fe/H] based on our program stars can be entirely attributed
to the measurement uncertainty. The mean metallicity and dispersion are
essentially unchanged if we exclude the ``neutron-capture poor'' star A124 with
[Fe/H] = $-$1.19. 

\begin{figure}
\centering
      \includegraphics[width=.99\hsize]{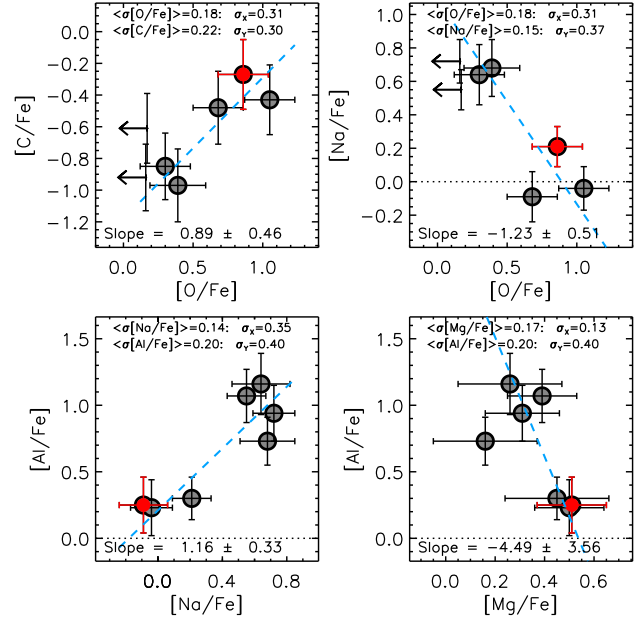} 
      \caption{Abundance ratios for combinations of the light elements (C, O,
Na, Mg and Al). The red data point is star A124, which has low neutron-capture
element abundance ratios compared to the remaining program stars. The dashed
line is the linear fit to the data (slope and error are included). The average
error ($<\sigma$[X/Fe]$>$) and dispersion ($\sigma$) in the x-direction and
y-direction are included.} 
         \label{fig:onamgalsi}
\end{figure}

For the light elements, we find evidence for a star-to-star variation for C, O,
Na and Al and perhaps Mg (see Figure \ref{fig:onamgalsi}). For O, Na and Al,
the observed abundance dispersion exceeds the average measurement error by a
factor of $\sim$2 indicating a genuine abundance spread for these elements.
These abundance dispersions are confirmed by the statistical significance of
the correlations between the following pairs of elements, $\ga$ 3-$\sigma$ for
[O/Fe] vs.\ [Na/Fe] and [Na/Fe] vs.\ [Al/Fe]. There is a suggestion that Mg and
Al are anticorrelated, a characteristic found in other, but not all, globular
clusters.  There is no evidence for an abundance variation for Si, nor for
correlations between Si and all other elements as seen in NGC 6752
\citep{yong13}. 

\begin{figure}
\centering
      \includegraphics[width=.99\hsize]{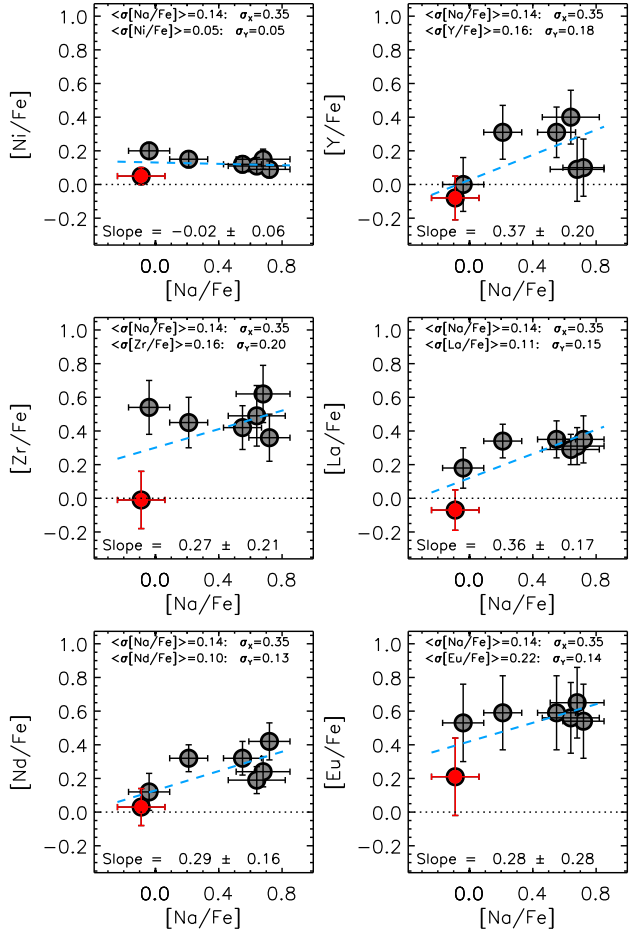} 
      \caption{Same as Figure \ref{fig:onamgalsi} but for Ni and the
neutron-capture elements versus Na.} 
         \label{fig:ncap}
\end{figure}

In Figure \ref{fig:ncap} we plot Ni and a selection of neutron-capture elements
versus Na.  Star A124, marked in red this and other figures, exhibits
systematically lower abundance ratios for the neutron-capture elements compared
to the remaining program stars. For a given element, we can compare the [X/Fe]
ratio for star A124 to the average [X/Fe] ratio for the other stars. When doing
this for the neutron-capture elements from Sr to Eu for A124, we find an
average difference of 0.29 $\pm$ 0.02 ($\sigma$ = 0.07). This difference is
approximately twice the average error and suggests that this star is chemically
different from the other six program stars, at least for the neutron-capture
elements. Thus, our quantitative analysis confirms the visual impression from
Figure \ref{fig:spec} in which star A124 has weaker Zr lines relative to other
program stars of comparable stellar parameters. Aside from star A124, we find
no compelling evidence for an abundance dispersion for the neutron-capture
elements. In all cases, the observed abundance dispersion is consistent with
the measurement uncertainty. We find no evidence for a statistically
significant correlation between any of the heavy elements and Na; such a
correlation has been identified in the globular cluster NGC 6752
\citep{yong13}. 

\begin{figure}
\centering
      \includegraphics[width=.65\hsize]{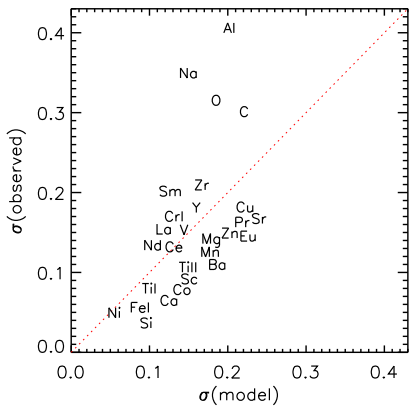} 
      \caption{Measured abundance dispersion, $\sigma$(observed, including star
A124) versus average measurement error, $\sigma$(model). The dotted red line is
the 1:1 relation.  \feii\ is not included (see text for details). For clarity,
some values have been shifted by up to 0.01 dex.} 
         \label{fig:errors}
\end{figure}

In Figure \ref{fig:errors}, we compare the measured abundance dispersion
(including star A124) with the average total error. As already noted, the light
elements O, Na and Al exhibit significantly larger dispersions than expected
given the measurement uncertainty. For all other elements, with the possible
exception of C, the observed abundance dispersions can be attributed to the
measurement uncertainty. Note that the abundances for \feii\ were forced to
match those of \fei, and this species is not included in Figure
\ref{fig:errors}. Had we excluded star A124 when generating this figure, the
$\sigma$(observed) values would have decreased such that a larger proportion of
the data would lie below the 1:1 relation. As such, our measurement
uncertainties might be overestimated. 

\begin{figure*}
\centering
      \includegraphics[width=.65\hsize]{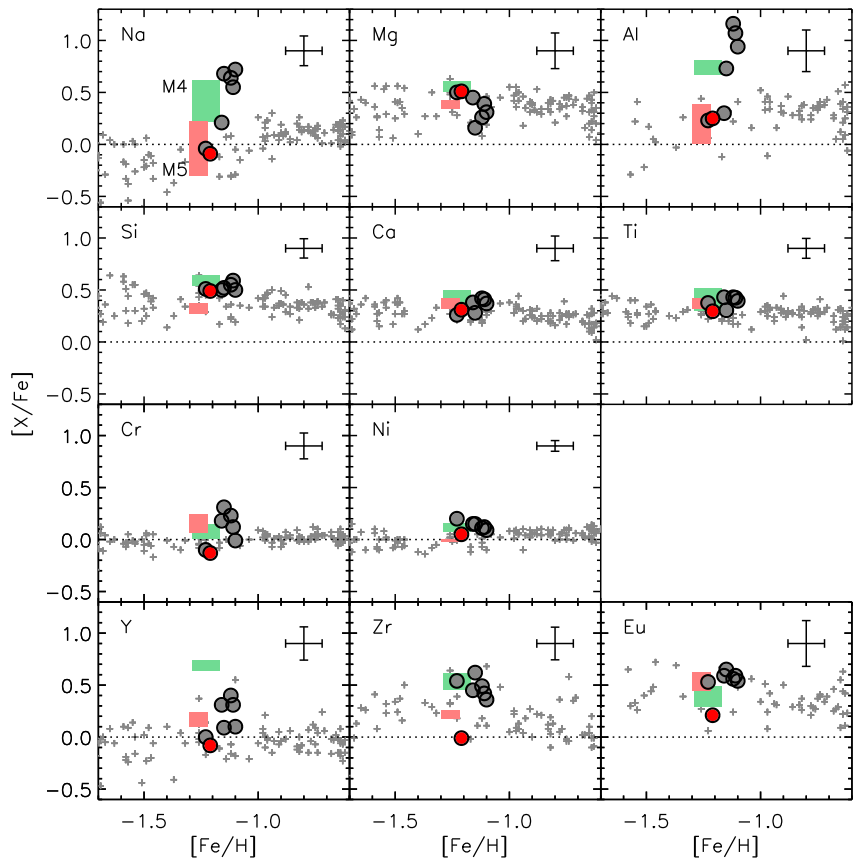}
      \caption{Abundance ratios [X/Fe] vs.\ [Fe/H] for the program stars (the
red circle is star A124 with low neutron-capture elements ratios). For Ti, we
adopt the average of \tii\ and \tiii. A representative error bar is included in
each panel. The small grey symbols are field halo stars taken from
\citet{fulbright00}. The shaded aqua and pink regions represent the data (mean
value $\pm$ standard deviation) for M4 and M5, respectively, taken from
\citet{yong08m4m5a,yong08m4m5b}.} 
         \label{fig:field}
\end{figure*}

In Figure \ref{fig:field}, we plot [X/Fe] vs.\ [Fe/H] for the program stars and
field stars from \citet{fulbright00}. For Na and Al, one can clearly see that
the abundance dispersions for [Na/Fe] and [Al/Fe] in M62 far exceed that of the
field stars of comparable metallicity and that the lower envelope of the
cluster values are in good agreement with field stars. Such results have been
well established for other clusters (e.g., see reviews by
\citealt{kraft94,gratton04}; \citealt{gratton12}). We note that star A124 has Na
and Al abundances in agreement with field stars. Further, the neutron-capture
element abundances for this star, while lower than the cluster average, appear
typical for field halo stars at the same metallicity, which is consistent with
it being a field interloper. 

For Si, Ca, Ti, Cr and Ni, the abundance ratios for M62 are in good agreement
with field stars of the same metallicity. In solar system material, elements
mainly produced via the $s$-process include Y and Zr, and we refer to these as
$s$-process elements. Similarly, elements predominantly produced via the
$r$-process in solar system material are referred to as $r$-process elements
and include Eu. For the $s$-process elements Y and Zr, the M62 values lie
slightly above field halo stars of comparable metallicity. For the $r$-process
element Eu, the M62 values lie near the upper envelope defined by the field
stars. (We caution that zero-point offsets may reduce, or amplify, the
abundance differences between M62 and the field stars.) 

In Figure \ref{fig:field}, we also include the comparable metallicity globular
clusters M4 ([Fe/H] = $-$1.08; \citealt{ivans99}) and M5 ([Fe/H] = $-$1.21;
\citealt{ivans01}) using the data from \citet{yong08m4m5a,yong08m4m5b}. For Na
and Al, while the abundance dispersions found in M4 and M5 are considerably
smaller than in M62, we note that our sample sizes are small. Examination of
the larger samples in \citet{ivans99,ivans01} and \citet{marino08} also
indicate that the abundance dispersions for Na and Al in M4 and M5 are probably
smaller than those of M62.  For the elements Si to Ni, the abundance ratios
[X/Fe] in M4 and M5 match those of M62 and field stars in their metallicity
range. For the $s$-process elements, M4 is known to have larger [X/Fe] ratios
than M5. 

We conclude this section by noting some physical parameters for M4, M5 and M62
which may provide insight into any chemical abundance differences between these
clusters.  \citet{deangeli05} measured relative ages for these three clusters
and found M5 to be 10-15\% younger\footnote{\citet{deangeli05} regard their
typical uncertainties to be 5\% to 8\%.} than M4 and M62 with the latter two
being coeval. \citet{marinfranch09} also found M5 to be slightly younger than
M4. On the other hand, \citet{vandenberg13} measured ages for M4 and M5 of
11.50 Gyr, but did not study M62. The absolute luminosities for M4, M5 and M62
are $-$7.19, $-$8.81 and $-$9.18, respectively, and the central velocity
dispersions are 4.0, 5.5 and 14.3 \kms, respectively \citep{harris96}.  M4 and
M62 have orbits that are restricted to the inner disk and bulge, apocentric
radii of $R_a$ = 5.9 kpc \citep{dinescu99} and $R_a$ = 3.1 kpc \citep{allen06},
respectively while M5 has an orbit which may be consistent with an accretion
event, $R_a$ = 35.4 kpc. 

\section{DISCUSSION}

The focus of this discussion is to examine the nature of M62 in light of its
chemical abundances. In Table \ref{tab:abun2}, we present the average cluster
abundance ratios with and without star A124, which has lower neutron-capture
element abundances.  

\begin{table}
 \centering
 \begin{minipage}{90mm}
  \caption{Mean abundances and abundance dispersions for the program stars
(excluding abundance limits).} 
  \label{tab:abun2} 
  \begin{tabular}{@{}lcrcrc@{}}
  \hline
	Species & 
	Z & 
	[X/Fe] & 
	$\sigma$[X/Fe]\footnote{Standard deviation} & 
	[X/Fe] & 
	$\sigma$[X/Fe]$^a$ \\ 
	& 
	& 
	\multicolumn{2}{c}{All stars} &
	\multicolumn{2}{c}{Excluding A124} \\
\hline 
O\,{\sc i}       &   8 &    0.52 &    0.35 &    0.49 &    0.38 \\ 
Na\,{\sc i}      &  11 &    0.38 &    0.35 &    0.46 &    0.31 \\ 
Mg\,{\sc i}      &  12 &    0.37 &    0.13 &    0.34 &    0.13 \\ 
Al\,{\sc i}      &  13 &    0.67 &    0.40 &    0.74 &    0.39 \\ 
Si\,{\sc i}      &  14 &    0.52 &    0.04 &    0.53 &    0.04 \\ 
Ca\,{\sc i}      &  20 &    0.35 &    0.06 &    0.35 &    0.07 \\ 
Sc\,{\sc ii}     &  21 &    0.20 &    0.09 &    0.23 &    0.06 \\ 
Ti\,{\sc i}      &  22 &    0.34 &    0.08 &    0.36 &    0.06 \\ 
Ti\,{\sc ii}     &  22 &    0.42 &    0.09 &    0.43 &    0.10 \\ 
V\,{\sc i}       &  23 &    0.22 &    0.15 &    0.27 &    0.11 \\ 
Cr\,{\sc i}      &  24 &    0.09 &    0.17 &    0.12 &    0.15 \\ 
Mn\,{\sc i}      &  25 & $-$0.30 &    0.12 & $-$0.30 &    0.13 \\ 
Fe\,{\sc i}\footnote{[\fei/H] or [\feii/H]} &  26 & $-$1.15 &    0.05 & $-$1.15 &    0.05 \\ 
Fe\,{\sc ii}$^b$ &  26 & $-$1.16 &    0.05 & $-$1.15 &    0.05 \\ 
Co\,{\sc i}      &  27 &    0.08 &    0.08 &    0.09 &    0.08 \\ 
Ni\,{\sc i}      &  28 &    0.12 &    0.05 &    0.14 &    0.04 \\ 
Cu\,{\sc i}      &  29 & $-$0.22 &    0.18 & $-$0.19 &    0.17 \\ 
Zn\,{\sc i}      &  30 &    0.11 &    0.15 &    0.10 &    0.16 \\ 
Rb\,{\sc i}\footnote{Measurements only for stars c53 and d10.} & 37 &     0.11 &    0.03 &    0.11 &    0.03 \\
Sr\,{\sc i}      &  38 & $-$0.39 &    0.17 & $-$0.33 &    0.08 \\ 
Y\,{\sc ii}      &  39 &    0.16 &    0.18 &    0.20 &    0.16 \\ 
Zr\,{\sc i}      &  40 &    0.41 &    0.20 &    0.48 &    0.09 \\ 
Ba\,{\sc ii}     &  56 &    0.32 &    0.11 &    0.36 &    0.03 \\ 
La\,{\sc ii}     &  57 &    0.25 &    0.15 &    0.30 &    0.07 \\ 
Ce\,{\sc ii}     &  58 & $-$0.01 &    0.13 &    0.04 &    0.07 \\ 
Pr\,{\sc ii}     &  59 &    0.18 &    0.15 &    0.23 &    0.09 \\ 
Nd\,{\sc ii}     &  60 &    0.23 &    0.13 &    0.27 &    0.11 \\ 
Sm\,{\sc ii}     &  62 &    0.32 &    0.20 &    0.37 &    0.14 \\ 
Eu\,{\sc ii}     &  63 &    0.52 &    0.14 &    0.58 &    0.04 \\ 
Pb\,{\sc i}$^c$  &  82 & $-$0.20 &  \ldots & $-$0.20 &  \ldots \\
\hline
\end{tabular}
\end{minipage}
\end{table}

\subsection{The light elements}

It is now well established that every globular cluster exhibits star-to-star
abundance variations for C, N, O, F, Na, Mg and Al (e.g., see reviews by
\citealt{kraft94,gratton12}). These abundance variations are believed to have
been produced via hydrogen-burning at high temperature in asymptotic giant
branch (AGB) stars, fast-rotating massive stars and/or massive binaries (e.g.,
\citealt{fenner04,ventura05,decressin07,demink09,marcolini09}). 

\citet{carretta06} and \citet{carretta10mv} found that the abundance
dispersions for the light elements correlate with various physical parameters.
In particular, they noted that the interquartile range (IQR) for [O/Fe],
[Na/Fe], [O/Na] and [Mg/Al] exhibits a correlation with the absolute
luminosity, i.e., total mass. Within our modest sample, assuming for the
purposes of this exercise that the [O/Fe] limits are detections, we find
IQR[O/Na] = 1.11 and IQR[Mg/Al] = 0.89. With $M_V$ = $-$9.18 \citep{harris96},
the IQR values for M62 are consistent with those of other luminous clusters. 

\subsection{The neutron-capture elements} 

\begin{figure*}
\centering
      \includegraphics[width=.75\hsize]{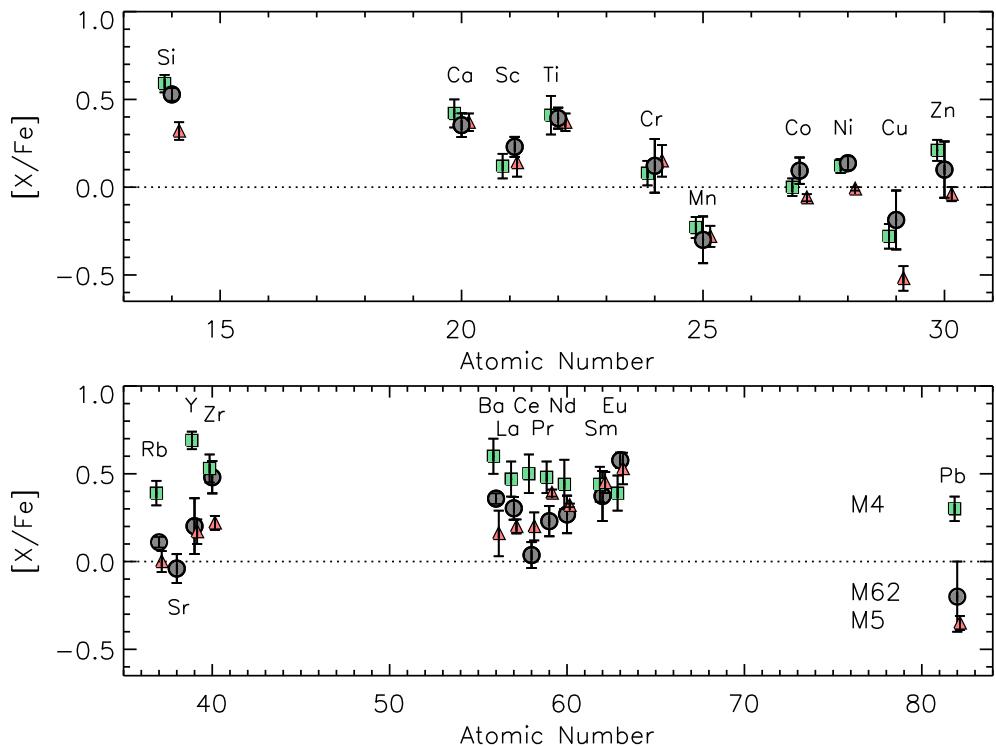}
      \caption{Abundance ratios [X/Fe] vs.\ atomic number for M62 (large grey
circles), M4 (small aqua squares) and M5 (small pink triangles). For M62, we
exclude star A124. For M4 and M5, the data are from \citet{yong08m4m5b}. For
Sr, we apply a non-LTE correction of 0.29 dex (see Section 4.2 for details) and
the M4 and M5 Sr values are not plotted since they are derived from a different
line than in the present work.} 
         \label{fig:m62m4m5}
\end{figure*}

In Figure \ref{fig:m62m4m5} we plot [X/Fe] versus atomic number for M62, M4 and
M5.  Regarding the $s$-process elements, Zr, Ba and La have ratios of [X/Fe]
$\simeq$ +0.4. Such values differ from the majority of field halo stars of
similar metallicity and from the comparable metallicity globular cluster M5. On
the other hand, these abundance ratios in M62 are comparable (but not
identical) to M4, a cluster believed to have formed from gas enriched in
$s$-process material.  It is intriguing that both M4 and M62 are inner disk
globular clusters of similar metallicity. Enhancements in the $s$-process
elements are also seen in the globular clusters $\omega$ Cen and M22
\citep[e.g.,][and references therein]{dacosta11}. For elements other than Zr,
Ba and La, the abundance pattern in M62 more closely resembles that of M5
rather than M4. 

At the metallicity of these clusters, published neutron-capture element
abundances in inner disk and bulge stars are limited. The \citet{fulbright00}
comparison field stars in Figure \ref{fig:field} are solar neighbourhood
objects. Since the pioneering work by \citet{mcwilliam94} on chemical
abundances in the bulge, the handful of papers presenting neutron-capture
element abundances at or below [Fe/H] = $-$1.0 in the inner disk or bulge
include \citet{johnson12} and \citet{bensby13}.  Although their sample sizes
are modest, some stars are enriched with material produced by $s$-process
nucleosynthesis.  Among the bulge and inner disk globular clusters with [Fe/H]
$\lesssim$ $-$1.0, NGC 6522 exhibits enhancements in the $s$-process elements
\citep{barbuy09,chiappini11} while HP-1 \citep{barbuy06} and NGC 6558
\citep{barbuy07} do not. 

The abundances of Rb and Pb offer important diagnostics regarding the nature of
the $s$-process. Due to a branching point at $^{85}$Kr along the $s$-process
path, the Rb/Zr ratio is sensitive to the neutron density at the site of the
$s$-process. At high neutron density, the [Rb/Zr] ratio is expected to be a
factor of $\sim$10 larger than at low neutron density
\citep{beer89,kappeler89}, although the yields depend on details in the models
and nuclear reaction rates \citep{pignatari08,pignatari10,vanraai12}. Indeed,
large Rb abundances have been observed in massive AGB stars
\citep{garciahernandez06,garciahernandez09}.  Pb and Bi are the last stable
nuclei along the $s$-process path. If the total neutron exposure, or neutron
supply per seed nuclei, is sufficiently high, large overabundances of these
elements are predicted \citep[e.g.,][]{gallino98,busso99,busso01,goriely00}, and
have been observed \citep[e.g.,][]{vaneck01,ivans05}. 

As already noted, measurements of Rb and Pb were only possible for the two
program stars observed at the Magellan Telescope. In Figure \ref{fig:rb} we
show observed and synthetic spectra near the 7800\AA\ \rbi\ feature. We measure
values of [Rb/Fe] = +0.09 and +0.13 for stars c53 and d10, respectively. As can
be seen in Figure \ref{fig:m62m4m5}, those values fall closer to M5 ([Rb/Fe] =
0.00) than to M4 ([Rb/Fe] = +0.39; \citealt{yong08m4m5b}). \citet{dorazi13}
measured [Rb/Fe] in three globular clusters (NGC 6752, NGC 1904 and 47 Tuc) and
found constant values in each cluster at the [Rb/Fe] $\simeq$ 0.0 level. The
[Rb/Zr] ratios are $-$0.04 and $-$0.05 for stars c53 and d10, respectively,
when the Zr abundances are shifted on the \citet{smith00} scale as described in
\citet{yong06}. The [Rb/Zr] ratios in M62 lie below those of M4 and M5, [Rb/Zr]
= +0.17 and +0.08, respectively, but above the $\omega$ Cen values
\citep{smith00}. In summary, the low [Rb/Fe] and [Rb/Zr] ratios in M62 offer no
support for a weak $s$-process contribution (i.e., the $s$-process in massive
stars from the $^{22}$Ne($\alpha$,$n$)$^{25}$Mg reaction) from which the high
neutron density would result in considerably higher [Rb/Fe] and [Rb/Zr] ratios
than observed. We will revisit the [Rb/Zr] ratio in the context of
fast-rotating massive stars later in this subsection. 

The Sr abundance in M62, $<$[Sr/Fe]$>$ = $-$0.39, appears to be unusually low
compared to the neighbouring elements, $<$[Rb/Fe]$>$ = +0.11 and $<$[Y/Fe]$>$ =
+0.16.  Since our analysis only considered one line, \sri\ 4607.33\AA, we
regard the low Sr abundance with some caution. Indeed, other studies that
consider the 4607.33\AA\ \sri\ line in globular cluster giants find [Sr/Fe]
ratios that are lower than for the neighbouring elements [Rb/Fe], [Y/Fe] and
[Zr/Fe] \citep[e.g.,][Yong et al.\ 2014, in preparation]{roederer11}.
\citet{bergemann12} investigated non-LTE corrections for the 4607.33\AA\ Sr
line, and for \teff/\logg/[Fe/H] = 4400/2.2/$-$1.2, the correction is 0.29 dex.
The magnitude and sign of this correction would help to reconcile the Sr
abundances with those of Rb and Y. In Figure \ref{fig:m62m4m5} we adjust the Sr
abundance accordingly. 

In Figure \ref{fig:pb}, we plot observed and synthetic spectra near the
4058\AA\ \pbi\ feature. We measure values of [Pb/Fe] $<-$0.09 and [Pb/Fe] =
$-$0.20 for stars c53 and d10, respectively. For comparison, the average values
in M4 and M5 are [Pb/Fe] = +0.30 and $-$0.35, respectively \citep{yong08m4m5b}.
Thus, the M62 values for [Pb/Fe] lie closer to M5 than to M4 (see Figure
\ref{fig:m62m4m5}).  \citet{roederer10b} argue that the Pb/Eu ratio offers a
good diagnostic of the presence of $s$-process enriched material. For M62, the
Pb to Eu ratios are at the same level as the (more metal-poor) ``$r$-process
standard stars'' in \citet{roederer10b}. 

\begin{figure}
\centering
      \includegraphics[width=.98\hsize]{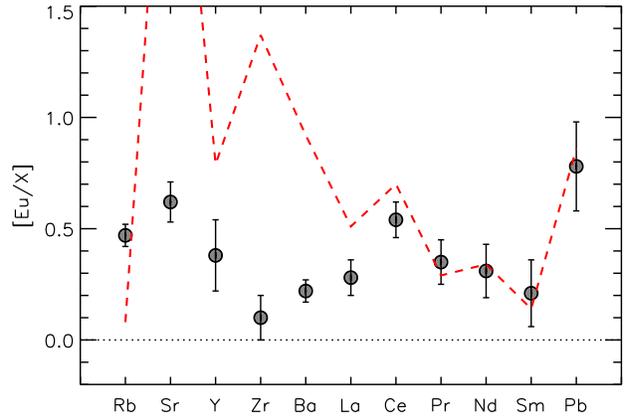}
      \caption{Abundance ratios [Eu/X] for X = Rb to Pb. The dashed red line
represents the scaled-solar $r$-process distribution \citep{bisterzo11}. (For
Sr, we apply a non-LTE correction of 0.29 dex, see Section 4.2 for details.)} 
         \label{fig:eux}
\end{figure}

From Table \ref{tab:abun2}, all element ratios [Eu/X] (for X = La and heavier
elements) are consistent, within $\sim$0.2 dex, with the scaled-solar
$r$-process distribution \citep{bisterzo11}, see Figure \ref{fig:eux}. (Note that
while we refer to $s$-process and $r$-process elements according to their 
origin in solar system material, those definitions may be incorrect in M62,
e.g., La and Ce in this cluster are likely produced primarily through the
$r$-process.) Regarding the lighter elements, the first to deviate from the
$r$-process signature is Ba: while the solar $r$-process value is [Eu/Ba]$_{\rm
r}$ $\sim$ 0.9 dex, M62 exhibits a ratio of $\sim$ 0.2 dex. In the same way,
all elements lighter than Ba depart from the solar system $r$-process
signature. 

The neutron-capture element abundance ratios of the M62 stars are not
consistent with $s$-process nucleosynthesis occurring in AGB stars of low
metallicity.  Low-mass AGB stars of $M \lesssim 3M_{\odot}$ are the main site
of the $s$-process in the Galaxy and typically produce [hs/ls]\footnote{$hs$
refers to the heavy $s$-process elements (Ba, La, etc.) and $ls$ refers to the
light $s$-process elements (Y, Zr, etc.).} $\approx 0.4$, [Pb/hs] $\ge 1.0$~dex
and [Rb/Zr] $\le 0$  at [Fe/H] $\approx -1.2$ (e.g.,
\citealt{busso01,cristallo11}; Fishlock et al.\ 2014, in preparation). This is
a direct consequence of the operation of the $^{13}$C($\alpha$,$n$)$^{16}$O
reaction which is the dominant neutron source for $M \lesssim 3M_{\odot}$
\citep[e.g.,][]{busso99,lugaro12,bisterzo12}. For intermediate-mass AGB stars
the $^{22}$Ne($\alpha$,$n$)$^{25}$Mg reaction is the neutron source and
operates during convective thermal pulses. The abundance pattern produced in
those stars should be dominated by lighter $s$-process elements (Y, Rb, Sr, Zr)
\citep{vanraai12,karakas12}. At [Fe/H] $\approx -1.2$, typical abundances in
stars $M \ga 3M_{\odot}$ are [hs/ls] $\approx -0.8$, [Pb/hs] $\approx -0.2$,
and [Rb/Zr] $\ga0.0$ dex (Fishlock et al. 2014, in preparation). 

Fast-rotating massive stars (which we hereafter refer to interchangeably as
``spin stars'') can produce a large amount of $s$-process products at low
metallicity, where the dominant neutron source is
$^{22}$Ne($\alpha$,$n$)$^{25}$Mg in the convective He-burning core and in the
subsequent convective C-burning shell \citep{pignatari08,frischknecht12}. In
these stars, $^{22}$Ne is primary, i.e., it is not directly produced from the
initial CNO abundances as in non-rotating massive stars (see for instance the
weak $s$-process in massive stars at solar metallicity, e.g.,
\citealt{the07,pignatari10}). The faster the star rotates, the more primary
$^{22}$Ne can be made available in the He core. The most abundant elements from
the $s$-process in spin stars are Sr, Y and Zr while production for heavier
elements up to Ba strongly depends on the rotation speed and on nuclear
uncertainties, e.g., on the $\alpha$-capture rates on $^{22}$Ne
\citep{pignatari08} and on the $\alpha$-capture rates on $^{17}$O (e.g.,
\citealt{best13}). The $s$-process in massive stars cannot produce elements
beyond the Ba neutron-magic peak (N = 82) due to basic properties of $^{22}$Ne as
a neutron source (e.g., \citealt{pignatari13}). 

Spin stars have been suggested as the source of the light element abundance
variations in globular clusters \citep{decressin07}.  Recall that in Figure
\ref{fig:ncap} there was no significant correlation between the abundances of
Na and the neutron-capture elements. Thus, there is no obvious connection
between the star-to-star Na abundance variation and the enhancement in Y, Zr
and Ba abundances in this cluster. 

\citet{pignatari08} explored the impact of changing the amount of primary
$^{22}$Ne available to make neutrons, from 1 percent (in mass fraction) to
0.2 percent, for a 25 \solmass\ model at [Fe/H] = $-$3. The result of this
change was an increase in the [Zr/Ba] ratio from $-$0.18 to 0.67. As expected,
Eu is not efficiently produced by the $s$-process regardless of the initial
abundance of $^{22}$Ne. M62 stars exhibits an average [Eu/Ba] = +0.2, and this
value lies between the $s$-process abundance signature of spin stars and the
$r$-process.  This intermediate value may be a consequence of the combination
of the $r$-process (which makes Eu and some Ba) and the $s$-process in spin
stars (producing Ba and only a small amount of Eu). 

M62 exhibits [Rb/Zr] and [Rb/Y] ratios of $\sim-$0.4 and $\sim-$0.1,
respectively.  In the models considered by \citet{pignatari08}, reducing the
initial $^{22}$Ne abundance from 1 per cent (in mass fraction) to 0.2 percent
increases the [Rb/Zr] ratio from $-$0.86 to $-$0.57, and the [Rb/Y] ratio from
$-$0.86 to $-$0.75.  The predicted [Rb/Zr] ratio is slightly lower than the
observations.  Those predictions were based on a 25 \solmass\ spin star model.
It would be of great interest to explore a larger range of initial masses in
order to quantify the final core-collapse supernova yields for the elements
from Rb to Zr. On the other hand, it is difficult to reconcile the average
[Rb/Y] observed in M62 with spin star nucleosynthesis. Y is efficiently
produced at high neutron densities from Sr.  The dispersion in the [Y/Fe] ratio
is consistent with the measurement uncertainties, although there is a hint that
the [Y/Fe] ratios cluster around two values, $\sim$0.1 and $\sim$0.4. 

The abundance enhancement at Y and Zr is a key feature of the lighter element
primary process (LEPP) exhibited by some halo stars at lower metallicities,
[Fe/H] $\la$ $-$2 (e.g., \citealt{travaglio04}). While the astrophysical site
of the LEPP is not clearly understood, a number of scenarios have been
proposed, including explosive nucleosynthesis triggered by neutrino-winds in
core-collapse supernovae (e.g.,
\citealt{woosley92,froehlich06,farouqi09,arcones11}) and the $s$-process in
fast-rotating massive stars (e.g., \citealt{pignatari08}). While the
aforementioned LEPP explosive scenarios cannot efficiently contribute to
elements up to the Ba peak, the $s$-process in spin stars may potentially
produce them. 

To summarise, while the abundance ratios for the elements Y, Zr and Ba may be
consistent with the $s$-process in spin stars, the high [Rb/Y] ratio is
difficult to explain.  Spin-star models at metallicities closer to that of M62
and for a larger range of initial masses would be needed to perform a
consistent chemical evolution study, and test the spin stars scenario for this
cluster. 

In this context, it has been reported that NGC 6522, a bulge globular cluster,
has enhanced Sr and Y abundances consistent with spin star nucleosynthesis
\citep{chiappini11}.  In that study, the abundances were derived from the \sri\
6503.99\AA\ and \yii\ 6613.73\AA\ lines. As noted in their paper, the \sri\
line is weak. We suspect that this \yii\ line is blended with the \fei\
6613.82\AA\ line \citep{nave94}, and this Fe line has a lower excitation
potential and $\log gf$ value of 1.011 eV and $-$5.587, respectively (Vienna
Atomic Line Database, \citealt{kupka99}). It would therefore be interesting to
re-examine the Sr and Y abundances in NGC 6522 using additional lines to
confirm the Sr and Y enhancements in that cluster. 

\subsection{The lack of abundance dispersions} 

Our analysis reveals no evidence for an intrinsic metallicity dispersion in
M62. As noted in the introduction, the globular clusters with metallicity
dispersions are preferentially the more luminous objects and tend to have
extended horizontal branches. Some, and perhaps all, of these clusters may be
the nuclei of accreted dwarf galaxies.  Based on our results, M62 is a massive
globular cluster with an extended horizontal branch that does not harbour a
metallicity dispersion. 

Star A124 has lower ratios of the neutron-capture elements when compared to the
other program stars. Although A124 has stellar parameters and a radial velocity
consistent with cluster membership, the Besan\c{c}on model \citep{robin03}
predicts that some field contamination is likely, $\sim$4\%. The simplest
explanation is that A124 is a field star. The O and Na abundances for A124 lie
at the lower end of the distribution defined by the cluster stars and are
similar to field stars. Proper-motion and/or parallax measurements will test
this hypothesis.  Nevertheless, future studies may reveal additional stars in
M62 with a range of neutron-capture element abundances. If M62 really hosts two
populations of stars, of equal size, with distinct neutron-capture element
abundances, then the probability of observing six out of seven program stars
from a single population is 6\%. 

\section{CONCLUDING REMARKS}

In this paper we present a chemical abundance analysis of seven bright giants
in M62, a luminous globular cluster with an extended horizontal branch. We find
no evidence for a metallicity dispersion, as seen in other luminous globular
clusters with similar horizontal branch morphology. We find star-to-star
abundance variations and correlations for O, Na and Al and the amplitudes of
those variations are comparable to those found in the most massive globular
clusters. The elements Zr, Ba and La exhibit enhancements, [X/Fe] $\sim$ +0.4,
compared to field stars at the same metallicity. The lack of a significant
correlation between the abundances of Na and the neutron-capture elements would
suggest that the light element variations and enhancement in Y, Zr and Ba are
produced in different processes and/or sites. 

For the elements heavier than La, the abundances are consistent with the
scaled-solar $r$-process distribution. On the other hand, Y, Zr and Ba are
clearly enhanced when compared to the scaled-solar $r$-process distribution.
The abundance pattern for these elements is incompatible with the s-process in
AGB stars or in non-rotating massive stars. On the other hand, while the
abundance distribution for these elements could be produced by fast-rotating
massive stars, the high [Rb/Y] ratios that we measure do not match existing
FRMS model predictions. 

One star has neutron-capture element abundance ratios that are distinct from
the remaining stars. While this is likely a field halo star, the identification
of additional stars in the vicinity of M62 with similar abundance patterns
would be of great interest. Studies of other luminous globular clusters are
necessary to complete the census and characterisation of the most luminous
clusters in the Milky Way. 

\section*{Acknowledgments}

We thank Christian Johnson for helpful discussions. 
D.Y, A.A.B, G.D.C, A.I.K and J.E.N gratefully acknowledge support from the
Australian Research Council (grants DP0984924, FS110200016, FT110100475 and
DP120101237). 
J.A.-G acknowledges support by the Chilean Ministry for the Economy,
Development, and Tourism's Programa Iniciativa Cient\'ifica Milenio through
grant P07-021-F, awarded to The Milky Way Millennium Nucleus; by Proyecto
Fondecyt Postdoctoral 3130552; by Proyecto Fondecyt Regular 1110326; and by
Proyecto Basal PFB-06/2007. 
M.P acknowledges support from the Ambizione grant of the SNSF (Switzerland) 
and EuroGenesis (MASCHE). 
Funding for the Stellar Astrophysics Centre is provided by The Danish National
Research Foundation. The research is supported by the ASTERISK project
(ASTERoseismic Investigations with SONG and Kepler) funded by the European
Research Council (Grant agreement no.: 267864).

\label{lastpage}

\end{document}